\documentclass[prd,twocolumn,amsmath,amsfonts,amssymb,showpacs]{revtex4} 
\usepackage{graphicx}
\usepackage{float}
\usepackage{subfig}
\usepackage{xcolor}
\def\beq{\begin{equation}}
\def\eeq{\end{equation}}
\def\bea{\begin{eqnarray}}
\def\eea{\end{eqnarray}}

\date{\today}

\begin{document}

 \vspace*{3cm}

\title{Quantum Thermodynamics and Quantum Entanglement Entropies in an Expanding Universe}

\author{Mehrnoosh Farahmand}
\affiliation{Department of Physics, University of Mohaghegh
Ardabili,  P.O. Box 179, Ardabil, Iran}

\author{Hosein Mohammadzadeh}
\affiliation{Department of Physics, University of Mohaghegh
Ardabili,  P.O. Box 179, Ardabil, Iran}

\author{Hossein Mehri-Dehnavi}
\affiliation{Department of Physics, Babol University of Technology,
Babol, 47148-71167, Iran}


\begin{abstract}
We investigate an asymptotically spatially flat Robertson-Walker spacetime from two different perspectives. First, using von Neumann entropy, we evaluate the entanglement generation due to the encoded information in spacetime. Then, we work out the entropy of particle creation based on the quantum thermodynamics of the scalar field on the underlying spacetime. We show that the general behavior of both entropies are the same. Therefore, the entanglement can be applied to the customary quantum thermodynamics of the universe. Also, using these entropies, we can recover some information about the parameters of spacetime.
\end{abstract}
\maketitle

\section{Introduction}
\label{sec:intrud}
To understand the nature of the universe, it is common to consider the curved and dynamical behavior of spacetime. In relativistic quantum field theory, the structure of spacetime affects the dynamic of fields. A great result of quantum field theory in curved spacetime is that the interaction between the quantum field and gravitational field gives rise to particle creation \cite{Davies}.

Recently, the entanglement of created particles in curved spacetime has been investigated extremely \cite{Benincasa,Mann1,Ebadi1,Ebadi2,Martinez2,Mohammadzadeh,Ball,Fuentes1,Moradi}. Also, some quantum information protocols has been considered in relativistic limit, specially in non-inertial frames \cite{Adesso,Alsing2,Helder,Mehri1,Alsing}.
The entanglement is observer dependent and is not invariant in dynamical spacetime. For scalar and spinor fields in dynamical background, one can recover some information about the nature of spacetime.
Investigation of created entanglement for arbitrary spin modes in expanding Robertson-Walker universe shows that the different behavior of scalar and spinor modes is related to the bosonic or fermionic properties \cite{Mohammadzadeh,Ball,Fuentes1}.
Recently, the teleportation between two observers in an asymptotically spatially flat Robertson-Walker expanding spacetime has been studied for bosons and fermions \cite{Helder}.

Thermodynamics is applied to description of small and large systems, such as black holes, dark matter and cosmology \cite{Zare,Ebadi3,Parker}. Recently, the thermodynamical properties of quantum fields in an expanding universe has been studied. It has been shown that one can define an inner friction due to quantum fluctuations of fields which has an entropic interpretation stemming from a quantum fluctuation relation \cite{Liu,Campisi}.
In other side, the entanglement was shown that can be applied in identifying the thermodynamic properties of the underlying spacetime structure \cite{Muller}. von Neumann entropy of generated entanglement is often employed to characterizing of the encoded information of spacetime \cite{Ball,Fuentes1}.
We will explore a relation between the entropy of quantum thermodynamics and the measure of generated entanglement.

We investigate entanglement and particle creation entropy of an expanding universe.
Particulary, entanglement and particle creation entropy are analyzed starting from vacuum in an asymptotically spatially flat Robertson-Walker spacetime using an exponential scale factor.
Based on the quantum information, we will extract the von Neumann entanglement entropy while based on the quantum thermodynamics of universe, we work out the particle creation entropy. The general behavior of the quantum information and quantum thermodynamics measures are compared. Also, we could apply particle creation entropy as a tool of determining the cosmological parameters of expanding universe with the exponential scale factor.

This paper is organized as follows: In section II we present the Klein-Gordon equation in curved spacetime for scalar fields and consider the asymptotically spatially flat spacetime in more details. In section III, we summarize the entanglement generation due to the expansion and evaluate it by von Neumann entropy. Section IV belongs to the evaluation of Bogoliubov coefficients of scalar fields in flat Robertson-Walker spacetime with exponential scale factor and the rate of particle creation. In section V, we work out the creation entropy based on the quantum thermodynamics of universe. We compare the well known entanglement entropy and the particle creation entropy in section VI. Finally, we conclude the paper in section VII.

\section{scalar field in expanding universe }
\label{sec:scalar}
Let us consider a $(1+1)$ dimensional expanding Robertson-Walker spacetime as follows
 \bea\label{metric}
   ds^{2}=\Omega^{2}(\eta)(d\eta^{2}-dx^{2}),
 \eea
where $\Omega$ is the scale factor and $\eta$ is the conformal time parameter, defined by relation $d\eta=a^{-1}dt$.
A time-like Killing vector field does not globally admit in such structure. 
 However, the metric, Eq. (\ref{metric}), is conformally flat, with a time-like conformal
 Killing vector. Provided that the conformal factor $\Omega(\eta)$ becomes
 constant as $\eta\rightarrow\pm\infty$ then a time-like Killing vector emerges
 asymptotically at past and future infinity, allowing the Klein-Gordon equation
 to have distinguishable positive and negative solutions.
 It is important to define particles that there exists such vector field. Only the observers flowing along time like vector field can describe particle states \cite{Davies}
 We now suppose a massive scalar field in Robertson-Walker spacetime, which is in analogue with inflationary cosmology \cite{Mukhanov}. The positive and negative frequency solutions of motion equation are recognized to define particle states. Thus the equation of motion of the scalar field is described by
\bea\label{eq1}
\left[\Box + \left(\frac{1}{6}+\xi \right)R + m^2 \right] \phi =0,
\eea
where the d'Alambertian operator is defined by $\Box\phi:=g^{\mu\nu}\nabla_\mu \nabla_\nu \phi=\frac{1}{\sqrt{-g}}\partial_\mu (\sqrt{-g} g^{\mu \nu} \partial_\nu \phi)$ in a curved spacetime. $R$ is the Ricci scalar and $\xi$ denotes the coupling between the field and gravity. The minimal and conformal coupling to gravity corresponds to $\xi=0$ and $\xi=\frac1 6$, respectively \cite{Davies}. We will set all constants $c,~\hbar,~G,~k_B$, equal to one, conventionally.

We assume the vacuum state of the field in the past infinity represents a thermal state from the perspective of an observer in the distant future. The quantization of the field is carried by its decomposition in terms of the positive and negative mode solutions of the Klein-Gordon equation, $u_k (\eta,x)$, as follows
\bea
\phi(\eta,x)=\int d^3k [ a_k u_k(\eta,x) + a_k^\dag u_k^*(\eta,x) ],
\eea
where $a_k,~a_k^\dag$ are annihilation and creation operators, respectively. $ (u_k , u_{k^\prime}) = \delta _{k {k^\prime}},~ (u_k^* , u_{k^\prime}^* ) = -\delta _{k {k^\prime}}, ~ (u_k , u_{k^\prime}^* ) = 0 $ are the orthonormality conditions. Because of the invariance with respect to the spatial translation, the solutions are separated into spatial and time components as follows
\bea\label{solution}
u_k=(2\pi)^{-\frac3 2} e^{ik.x} \chi_k(\eta).
\eea
Substituting the relation (\ref{solution}) in Eq. (\ref{eq1}), the equation of motion may be written as follows
\bea\label{motion Eq}
\frac {d^2\chi_k}{d\eta^2} + [ k^2 + \Omega^2 ( m^2 + \xi R ) ] \chi_k = 0.
\eea
We notice that $R(\eta)\longrightarrow0$ and $\Omega\longrightarrow$ ``constant" as $\eta\longrightarrow\pm\infty$, these conditions allow the definition of particle states in the asymptotic regions. Thus the pure positive frequency solution for $\eta\longrightarrow-\infty$ will be
\bea\label{in}
\chi_k (\eta) = (2\omega_k)^{- \frac 1 2} e^{-i\omega_k \eta},
\eea
where
\bea\label{omega}
\omega_k^2=k^2 + m^2 \Omega^2(\eta).
\eea
Eq. (\ref{in}) describe the particle states with respect to creation operators $a_k^\dag$.
For $\eta\longrightarrow+\infty$, only pure positive frequency solution will not generally remain. Therefore, the solution $u_k$ will be as the following
\bea
\tilde{\chi}_k (\eta) = (2\omega_k)^{- \frac 1 2} [ \alpha _k e^{-i\omega_k \eta } + \beta _k e^{+i\omega_k \eta} ].
\eea
$\alpha_k$ and $\beta_k$ are Bogoliubov coefficients. Using Eq. (\ref{in}), it can be rewritten such as follow
\bea\label{out}
\tilde{\chi_k} (\eta)= \alpha _k \chi _k  (\eta) + \beta _k \chi^* _k(\eta).
\eea
Therefore, in the infinite future the scalar field can be corresponding to a linear combination of positive and negative modes. It means that we could evaluate the annihilation (creation) operator in far past as a linear combination of annihilation (creation) operator in far future,
\bea\label{ak}
a_k = \alpha_k^* \tilde{a}_k - \beta_k^* \tilde{a}_{-k} ^\dag.
\eea
Using the canonical commutation relations for annihilation and creation operators, we obtain
\bea\label{ab}
|\alpha_k|^2 - | \beta_k|^2 =1.
\eea
If the system is initially defined in the vacuum state and there are no particles, the state of the system is given by $a_k |0_k\rangle =0 $ as follows $\eta\longrightarrow-\infty$.
Then the initial state converts to an exited state as $\eta\longrightarrow +\infty$ and the number of particles with respect to the $k$th mode is found as
\bea
\langle \tilde{N}_k\rangle = \langle0_k| \tilde{a}_k^\dag \tilde{a}_k |0_k\rangle,
\eea
therefore, we have
\bea\label{N}
\langle n_{\rm cr}\rangle = |\beta_k|^2,
\eea
where we have defined $\langle \tilde{N}_k\rangle \equiv \langle n_{\rm cr}\rangle$. As the conclusion, the ``in" vacuum (far past) regime corresponds to the excited states in ``out" vacuum regime.

\section{entanglement of created particles }
\label{sec:entangl}
The new vacuum state in distant past is a pure separable state while it is viewed as a pure entangled state from the perspective of distant future observer. In fact, we can show that the $k$th mode of ``in" vacuum, $|0_k\rangle\equiv|0_k 0_{-k}\rangle$, can be written as a Schmidt decomposition of ``out" state, $|\tilde{n}_k \tilde{n}_{-k}\rangle$. The distribution of Schmidt coefficients, $\lambda_i$, determines the amount of entanglement \cite{vedral}. If one of the coefficients is equal to one and others are zero, the ``in" and ``out" states are the same and there is no entanglement. If all of the Schmidt coefficients are equal, the state is maximally entangled. Therefore, we require a continuous function based on the conditions as follows,
\begin{equation}\label{con}
\left\{ {\begin{array}{*{20}{c}}
&&\hspace{ -.8cm}{S(\{ {\lambda _i}\} ) = 0,~~~~~~~~~{\lambda _i} = 1,~~~~~~{\lambda _{j \ne i}} = 0,}\\
&&\hspace{ -.6cm}{S(\{ {\lambda _i}\} ) = \log (d),~~~{\lambda _i} = \frac{1}{{\sqrt d }},~~~i = 1,...,d,}
\end{array}} \right.
\end{equation}
where $d$ is the minimum dimension of the Hilbert spaces of the systems. These are precisely the conditions which is considered to define the entropy. Therefore, the entanglement measure of two systems that form a pure state is given by the von Neumann entropy of (either of) the reduced density matrices.
\bea\label{Svon}
S(\rho_k) = - \sum |\lambda_i|^2 \log |\lambda_i|^2 = - {\rm Tr}[\rho_k \log (\rho_k)],
\eea
where
\bea\label{tr}
\rho_k = {\rm Tr}_{-k}[~\rho_k~] ={\rm Tr}_{-k}[~ |0_k\rangle \langle 0_k|~].
\eea

In previous section, we assumed that the system was in vacuum state and there was no entanglement in the remote past. The overall vacuum is defined by $|0_k\rangle = \mathop  \bigotimes \limits_{k =  - \infty }^\infty  |{\left. {{0_k}} \right\rangle}$. In the remote future, the entanglement will be generated. We restrict ourself to the modes of frequencies $k$ and $-k$ to find the relation between past and future states. The vacuum state in ``in" region as a two-mode squeezed state can be written as follows
\bea
|0_k 0_{-k}\rangle = \sum\limits_n A _n |\tilde{n}_k\rangle |\tilde{n}_{-k}\rangle.
\eea
We note that $\{|n_k\rangle\}$ and $\{|\tilde{n}_k\rangle\}$ represent the Hilbert-Fock states of the ``in" and ``out" regions, respectively.
Using definition of the vacuum state, i. e. $a_k|0\rangle=0$, and Eq. (\ref{ak}), we achieve the following relation
\bea
A_n = \left(\frac {\beta^*_k}{\alpha^*_k}\right)^n A_0.
\eea
Imposing the normalization condition $\langle0_k|0_k\rangle=1$, leads to
\bea
\left|A_0\right|^2= 1-\left|\frac {\beta_k}{\alpha_k}\right|^2,
\eea
and we obtain the vacuum state in the far past as follows
\bea\label{0k}
|0_k\rangle = \frac {1}{|\alpha_k|} \sum\limits_{n=0}^\infty \left(\frac {\beta^*_k}{\alpha^*_k}\right)^n |\tilde{n}_k\rangle |\tilde{n}_{-k}\rangle.
\eea
Using the von Neumann entropy we quantify the entanglement due to left $k$ and right $-k$ moving modes in the asymptotic future. To do this, we calculate the reduced density matrix by tracing out one of the two modes. According to Eq. (\ref{tr}) we obtain
\bea\label{ro}
\rho_k = \mathop {\rm Tr}\limits_{ - k} [\rho_0] = \frac {1}{|\alpha_k|^2} \sum\limits_{n} \left(\frac {|\beta_k|}{|\alpha_k|}\right)^{2n} |\tilde{n}_k\rangle \langle \tilde{n}_k|.
\eea
\\
The eigenvalues of $\rho_k$ are $(1-\gamma) \gamma^{n}$, where $\gamma\equiv|\frac {\beta_k}{\alpha_k}|^2$. Using Eq. (\ref{Svon}) the von Neumann entropy, entanglement entropy, is evaluated as follows
\bea\label{en}
S_{\rm en} = S(\rho_k) = \log \left( \frac {\gamma^\frac{\gamma}{\gamma-1}}{1-\gamma}\right).
\eea
\section{ a spacetime with exponential evolution and particle creation}
\label{sec:exponent}
There are different asymptotically spatially flat scale factors for Robertson-Walker spacetime in literature \cite{Birrell,Davies}. We consider an exponential scale factor as follows
\bea\label{scale}
\Omega^2(\eta)= c ~e^{-a |\eta|} + b^2,
\eea
where $a,~b,$ and $c$ are cosmological constant.
$\Omega(\eta)$ is not differentiable with respect to $\eta$ and it seems that such scale factor is unphysical, we note that the various components of Ricci curvature tensor are in fact continuous. We use this scale factor as it easily handled.
Also, the exponential scale factor represents a toy model which can be described the inflationary models \cite{Kallosh}.
By choosing the scale factor (\ref{scale}), in the limit $\eta\rightarrow\pm\infty$, the metric (\ref{metric}) will be as
$ds^{2}=b^{2}(\eta)(d\eta^{2}-dx^{2})$.
 Therefore, the metric is flat asymptotically to the past and future, and a time-like Killing vector is obtained in these limits.
We work by minimal coupling to gravity, i. e. take the case $\xi = 0$. For $\eta <0$, using Eq. (\ref{in}) and the normalization condition , $\chi_k \partial_\eta \chi_k^* - \chi_k^* \partial_\eta \chi_k = i$, we obtain
\bea\label{chi1}
\chi_k(\eta)= (2 \omega_k)^{-\frac1 2} \Gamma(1+\nu) (\frac1 2 \mu)^{-\nu} J_ \nu (\mu e ^{\frac1 2 \alpha \eta}),
\eea
where $J_\nu (x)$ are the Bessel functions of the first kind, $\nu = \frac{-2i\omega_k}{a}$, $\mu = \frac{2m c^ \frac1 2}{a}$ and $\omega_k(\eta\longrightarrow\infty) = \pm(k^2 + m^2 b^2)^\frac1 2$.
for $\eta>0$ the function (\ref{chi1}) is not a solution of Eq. (\ref{motion Eq}), due to absolute value in Eq. (\ref{scale}).
 Thus, for $\eta>0$ the general solution is found as follows
\bea\label{chi2}
\tilde{\chi_k}(\eta)= N J_ \nu (\mu e ^{-\frac1 2 \alpha \eta})+ M J_ {-\nu} (\mu e ^{-\frac1 2 \alpha \eta}),
\eea
where the constant $M$ and $N$ can be derived by matching the Eq. (\ref{chi1}) and Eq. (\ref{chi2}) and first derivatives in $\eta=0$. We note that at $\eta\longrightarrow+\infty$,
\bea\label{chi3}
\tilde{\chi_k}(\eta\longrightarrow+\infty)= N \frac{(\frac1 2 \mu)^{\nu}}{\Gamma (1+\nu)} e^{i\omega_k \eta} +  M \frac{(\frac1 2 \mu)^{\nu}}{\Gamma (1-\nu)} e^{-i\omega_k \eta}.\nonumber
\\
\eea
Now, using Eqs. Eq. (\ref{chi1}), Eq. (\ref{chi3}) and Eq. (\ref{out}) we extract the Bogoliubov coefficients as follows
\begin{widetext}
\bea
\alpha_k = 2(\frac1 2)^{-2\nu} \frac{\Gamma (1+\nu)}{\Gamma (1-\nu)}
\frac{[\nu_\mu^{-1}J_\nu(\mu)- J_{1+\nu} (\mu)]~ J_\nu (\mu)}{[J_{1-\nu}(\mu) J_\nu (\mu) + 2 \nu \mu^{-1} J_{-\nu}(\mu) J_\nu (\mu) - J_{-\nu}(\mu) J_{1+\nu}(\mu)]},\nonumber
\eea
\bea
\beta_k = \frac{J_{1-\nu}(\mu) J_\nu (\mu) + J_{-\nu}(\mu) J_{1+\nu}(\mu)}{J_{1-\nu}(\mu) J_\nu (\mu) + 2 \nu \mu^{-1} J_{-\nu}(\mu) J_\nu (\mu) - J_{-\nu}(\mu)J_{1+\nu}(\mu)}.\nonumber
\eea
\end{widetext}
Utilizing the properties of Bessel functions, Bogoliubov coefficients are determined as follows
\bea
&&\hspace{-.2 cm}\alpha_k = 2 \pi (\frac\mu 2 )^{1-2\nu} \csc(\pi \nu) \Gamma(1+\nu) J^\prime_\nu (\mu) J_\nu (\mu) / \Gamma(1-\nu),\nonumber
\eea
\bea
\beta_k = 1- \pi \mu \csc(\pi \nu)J^\prime_\nu (\mu) J_{-\nu} (\mu),\nonumber
\eea
where the primes indicates differentiation with respect to $\mu$. Then we have
\bea\label{alpha2}
|\alpha_k|^2 = \pi^2 \mu^2 \csc {\rm h}^2 (\pi |\nu|) |J^\prime_\nu (\mu) J_{-\nu} (\mu)|^2,
\eea
and
\bea\label{beta2}
|\beta_k|^2  = \pi^2 \mu^2 \csc {\rm h}^2 (\pi |\nu|) |J^\prime_\nu (\mu) J_{-\nu} (\mu)|^2 -1.
\eea
Therefore, from Eq. (\ref{N}) the expectation value of the number of particles in the mode $k$ is obtained such as follow
\bea\label{ncr}
\langle n_{\rm cr}\rangle  = \pi^2 \mu^2 \csc {\rm h}^2 (\pi |\nu|) |J^\prime_\nu (\mu) J_{-\nu} (\mu)|^2 -1.
\eea
\begin{figure*}[!ht]
  \begin{minipage}[t]{.48\textwidth}
  \center
  \includegraphics[width=1\columnwidth]{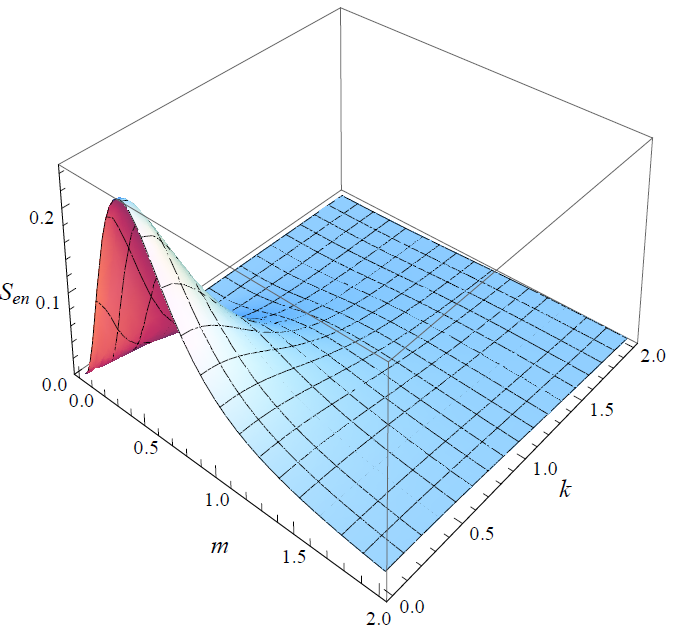}\\
  \caption{Entanglement entropy, $S_{\rm en}$, as a function of momentum $k$ and mass $m$ for fixed values $a=b=c=1$.
  }\label{Fig1}
\end{minipage}
  \begin{minipage}[t]{.48\textwidth}
  \center
  \includegraphics[width=1\columnwidth]{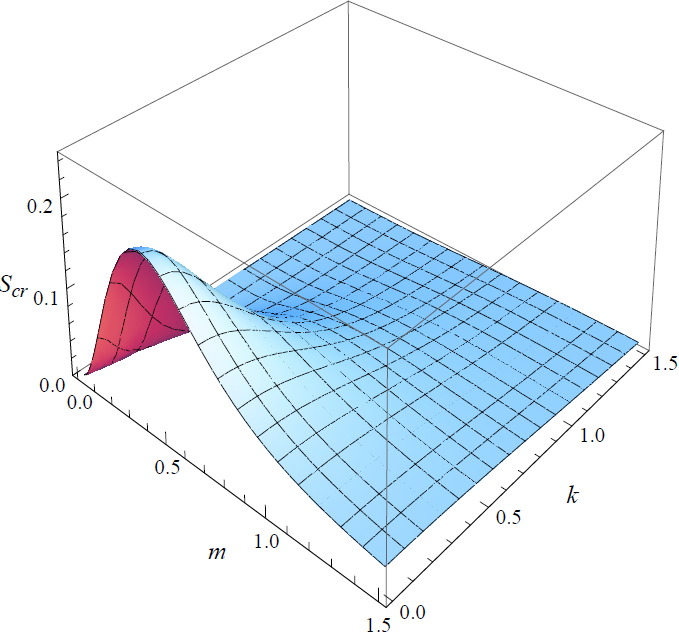}\\
  \caption{Particle creation entropy, $S_{\rm cr}$, as a function of momentum $k$ and mass $m$ for fixed values $a=b=c=1$.
  }\label{Fig2}
  \end{minipage}
\end{figure*}
\section{particle creation entropy}
\label{sec:entropy}
In the inflationary scenario which the universe expands very rapidly, each pair of modes undergoes a unitary evolution. Then we can independently investigate each pair of modes. This causes the mode pairs and external spacetime constituting a closed system as spacetime is expanding. Therefore,
We consider only interaction between the modes $k$ and $-k$, neglecting effects of the other modes of the field. Thus time evolution is unitary for each mode pair $(k,-k)$ and is modeled by two-mode squeezing.
This has a strongly resemblance with the quantum optical system by employing two-mode squeezing state that is under a classical external source \cite{Yurke,Alsing3,Gerry}. The classical background takes field away from equilibrium during expanding universe.
The operator $a_{-k}$ will be rewritten as $a_{-k}\longrightarrow b$ as a new notation. Thus the asymptotic past and future Hamiltonians are given by
\bea\label{H}
&& H= \omega_k (a_k ^\dag a_k + b_k ^\dag b_k +1),\\ \nonumber
&&\tilde{H} = \tilde{\omega}_k (\tilde{a_k}^\dag \tilde{a_k} + \tilde{b_k}^\dag \tilde{b_k}+1),
\eea
respectively.
The thermodynamical  frame work is introduced by the pair of field modes as a system, and the spacetime acts on the quantum field as a purely classical work reservoir. Therefore, we argue that the expansion of spacetime can be described as an unitary evolution on modes of opposite momenta. As an interesting view point, the expansion or changing the spacetime is doing work onto the quantum field.

The work has been done by spacetime which takes the field away from equilibrium, the average work, is obtained
 \bea
 W\equiv {\rm Tr} [\tilde{H}\rho_k] - {\rm Tr} [H \rho_k],
 \eea
where the first terms in right hand side denotes the average energy after and before evolution, respectively. If the system was initially in the vacuum state of remote past, $\rho_k =|0_k 0_{-k}\rangle \langle0_k 0_{-k}|$, using Eqs. (\ref{H}) we obtain
\begin{equation}\label{con}
\left\{ {\begin{array}{*{20}{c}}
&&\hspace{ -2.3cm}{{\rm Tr} [H \rho_k] = \omega_k},\\
&&\hspace{ -.6cm}{{\rm Tr} [\tilde{H} \rho_k] = (\langle n_{\rm cr}\rangle+1)~\tilde{\omega_k}},
\end{array}} \right.
\end{equation}
 and the average work which is done by spacetime will be such as follow
\bea\label{w}
\langle W\rangle = \tilde{\omega}_k ~ \langle n_{\rm cr}\rangle +(\tilde{\omega}_k -\omega_k).
\eea
The first term of right hand side is related to the energy of created particles while the second one shows the change in ground state energy. For a quantum adiabatic expansion of universe, there is not any particle creation and hence the second term is equal to adiabatic work from the total work.
\bea\label{wfric}
&&\langle W\rangle_{\rm adiabatic} = ( \tilde{\omega}_k - \omega_k),\\ \nonumber
&&\langle W\rangle_{\rm friction} = \langle W\rangle - \langle W\rangle_{\rm adiabatic} = \tilde{\omega}_k ~\langle n_{\rm cr}\rangle.
\eea
According to Eqs. (\ref{omega}) and (\ref{N}) for an exponentially expanding universe, we will have
\bea
&&\langle W\rangle_{\rm adiabatic} = 0,\\ \nonumber
&&\langle W\rangle_{\rm friction} = \langle W\rangle - \langle W\rangle_{\rm adiabatic} = \tilde{\omega}_k |\beta_k|^2.
\eea
N. Liu et al showed that how inner friction $\langle W\rangle_{\rm friction}$ can be interpreted as an entropy production in the cosmological context during particle creation \cite{Liu}. They used the fluctuation theorem \cite{Tasaki,Evans,Crooks,Campisi} and argued that the entropy production, $\langle s\rangle$, is such as follow
\bea\label{entropy}
\langle s\rangle=\frac{{\langle W\rangle}_{\rm friction}}{\tilde{T}} = \frac{\tilde{\omega}_k}{\tilde{T}} \langle n_{\rm cr}\rangle.
\eea
\begin{figure*}
  \begin{minipage}[t]{.48\textwidth}
  \center
   \includegraphics[width=0.95\columnwidth]{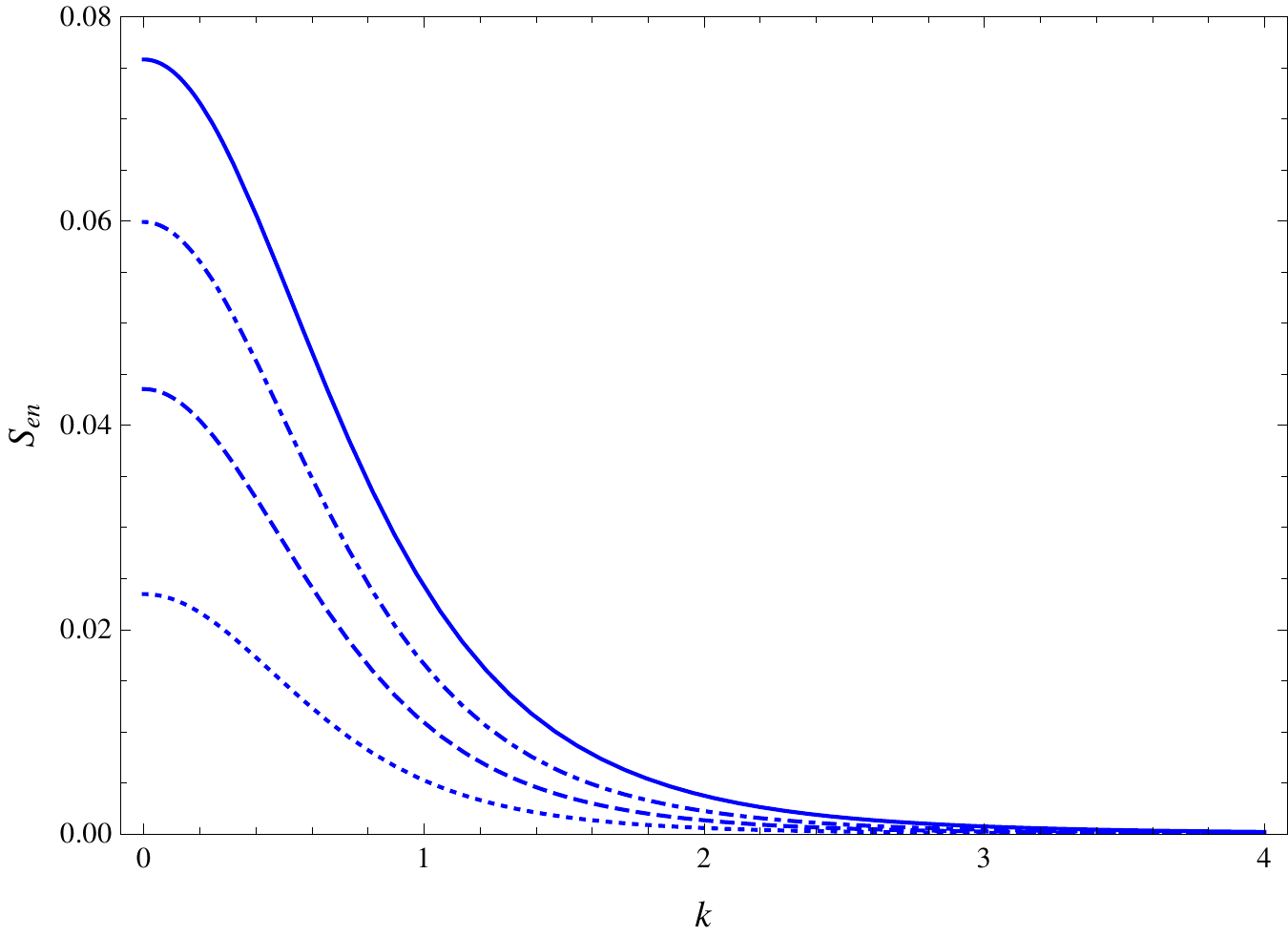}\\
  \caption{Entanglement entropy, $S_{\rm en}$, as a function of momentum $k$ for fixed values $m=a=b=1$ and the different
values of parameter $c$ that are $c= 1$ (solid), $c= 0.7$ (dotdashed),
$c= 0.5$ (dashed), and $c= 0.3$ (dotted) curves.
  }\label{Fig3}
\end{minipage}
  \begin{minipage}[t]{.48\textwidth}
  \center
 \includegraphics[width=0.95\columnwidth]{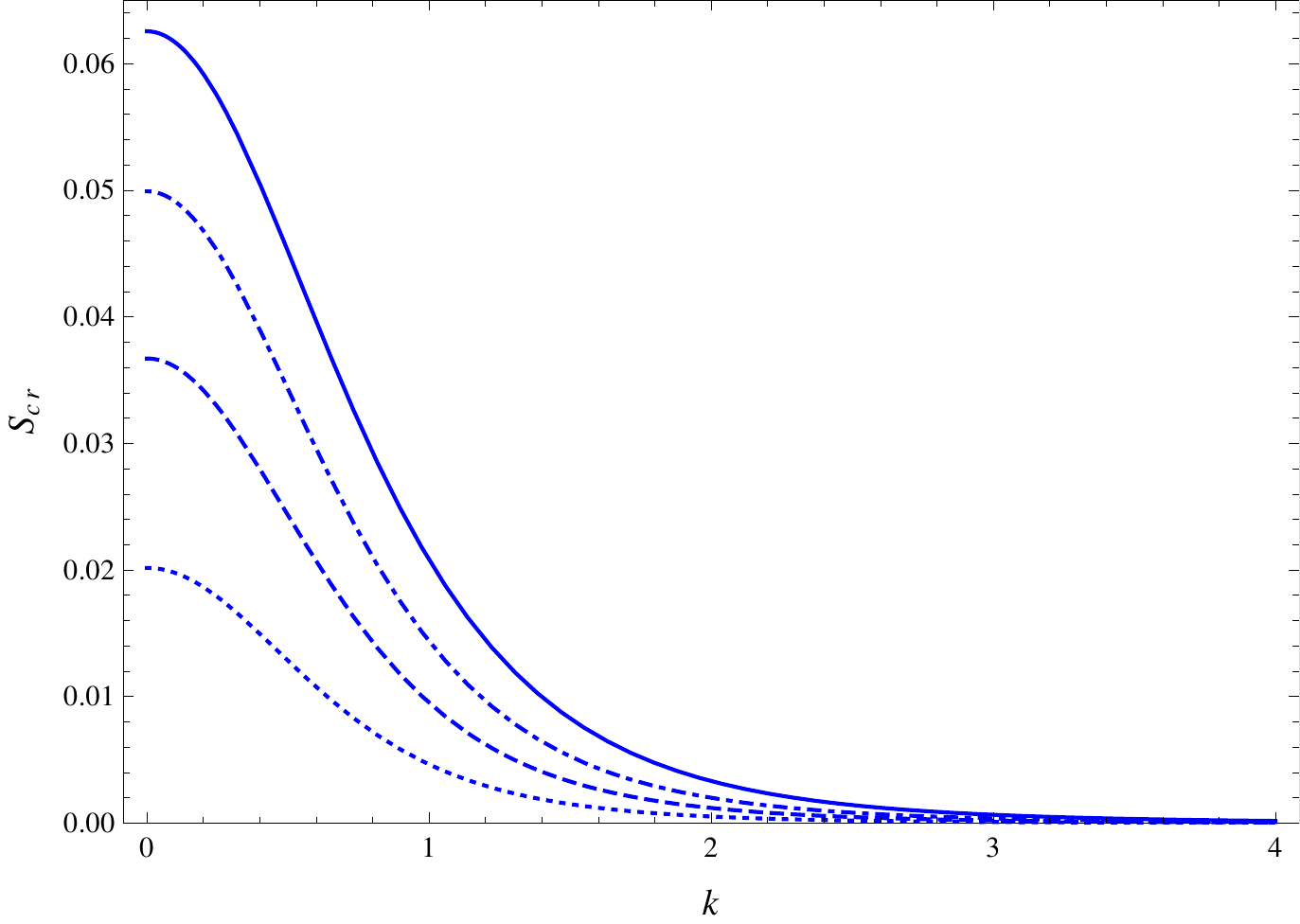}\\
  \caption{Particle creation entropy, $S_{\rm cr}$, as a function of momentum $k$ for fixed values $m=a=b=1$ and the different
 values of parameter $c$ that are $c= 1$ (solid), $c= 0.7$ (dotdashed),
$c= 0.5$ (dashed), and $c= 0.3$ (dotted) curves. }\label{Fig4}
\end{minipage}
    \begin{minipage}[t]{.48\textwidth}
  \center
  \includegraphics[width=0.95\columnwidth]{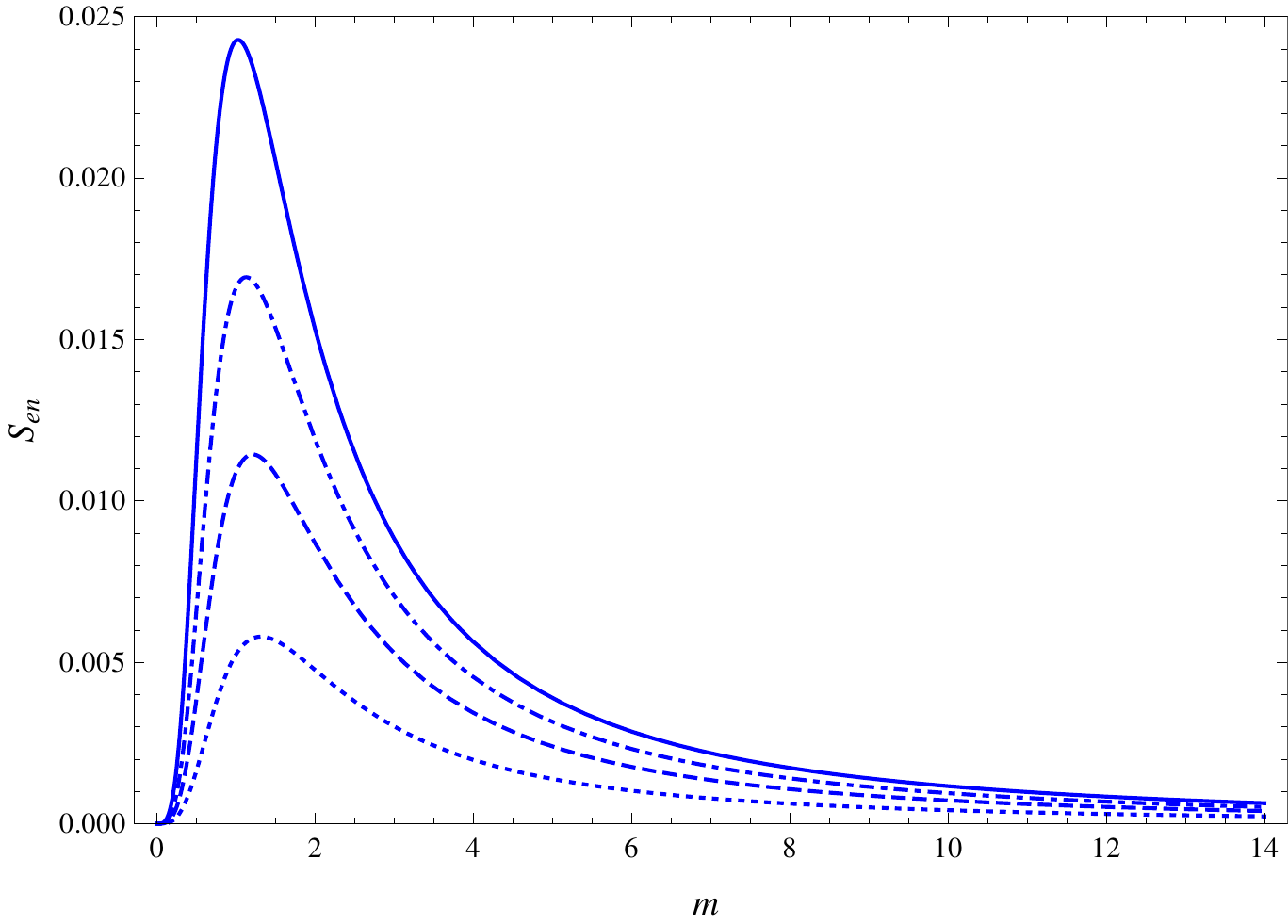}\\
  \caption{Entanglement entropy, $S_{\rm en}$, as a function of mass $m$ for fixed values $k=a=b=1$ and the different
values of parameter $c$ that are $c= 1$ (solid), $c= 0.7$ (dotdashed),
$c= 0.5$ (dashed), and $c= 0.3$ (dotted) curves. }\label{Fig5}
\end{minipage}
  \begin{minipage}[t]{.48\textwidth}
  \center
 \includegraphics[width=0.95\columnwidth]{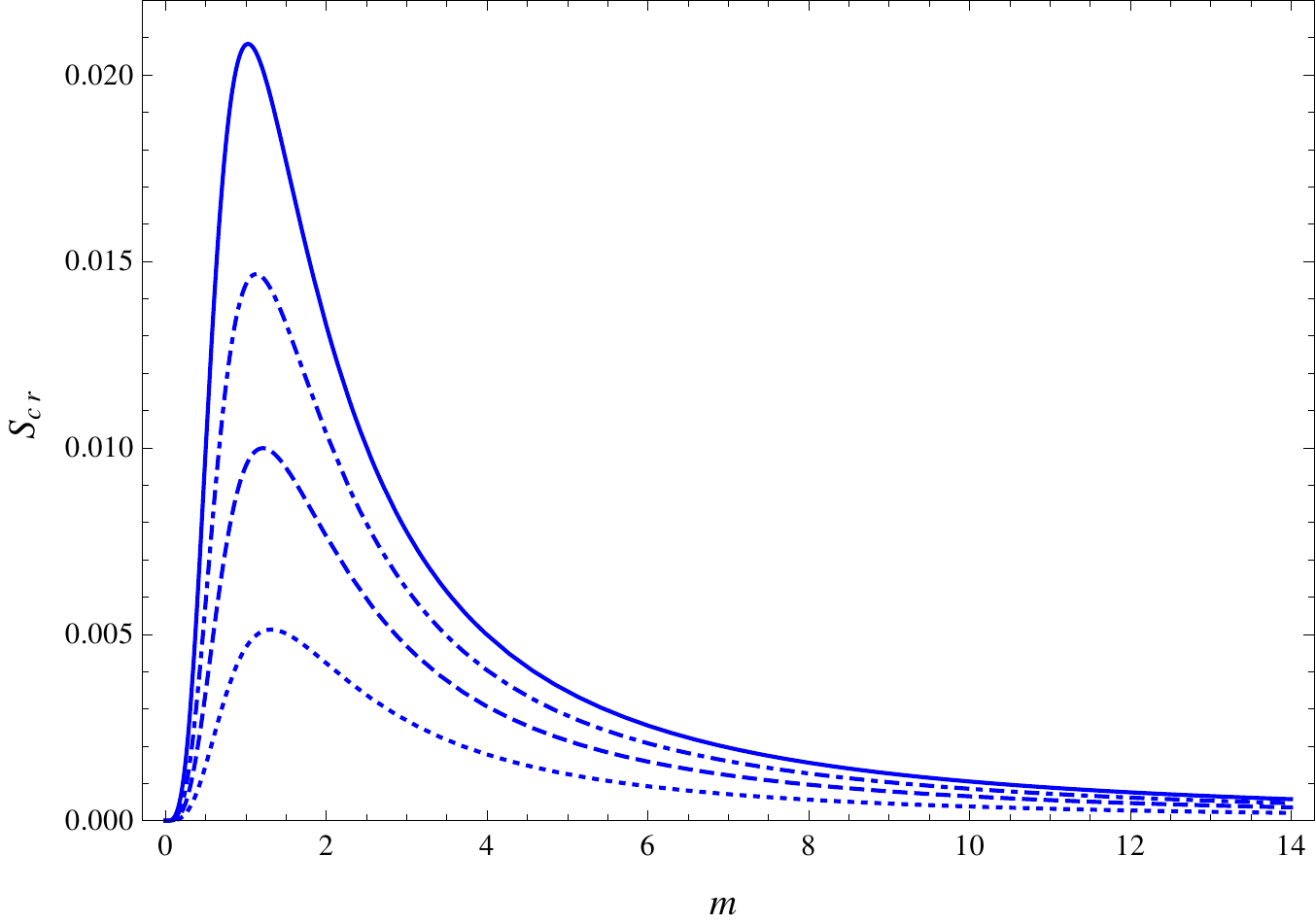}\\
  \caption{Particle creation entropy, $S_{\rm cr}$, as a function of mass $m$ for fixed values $k=a=b=1$ and the different
values of parameter $c$ that are $c= 1$ (solid), $c= 0.7$ (dotdashed),
$c= 0.5$ (dashed), and $c= 0.3$ (dotted) curves.}\label{Fig6}
  \end{minipage}
\end{figure*}

Using Eq. (\ref{ro}), we can extract the temperature $\tilde{T}$. we show that the reduced density matrix, $\rho_k$, is a thermalized state,
the Bogoliubov coefficient in Eq. (\ref{ab}) can be written as $|\alpha_k|\equiv\cosh z_k$ and $|\beta_k|\equiv\sinh z_k$ where $z_k$ is the squeezing parameter \cite{Barnett}.
 It is straightforward to show that Eq. (\ref{ro}) is written in the following form
\bea\label{rho}
\rho_k = 2 \sinh \frac{\tilde{\omega}_k}{2\tilde{T}} \sum\limits_{n_k} e^{-\frac{\tilde{\omega}_k}{\tilde{T}}(\tilde{n_k}+ \frac{1}{2})}|\tilde{n_k}\rangle\langle \tilde{n_k}|,
\eea
where $\tanh z_k \equiv e^{-\frac{\tilde{\omega}_k}{\tilde{T}}}$
 and $\frac{\tilde{\omega}_k}{2}$ is related to the ground state energy of the positive modes in Eq. (\ref{0k}). It is clear that the state Eq. (\ref{rho}) is a thermal state as follows
\bea\label{Gibbs}
\rho_k=(1-e^{-\frac{\tilde{\omega}_k}{\tilde{T}}})\sum\limits_{n_k} e^{-\frac{\tilde{\omega}_k}{\tilde{T}}\tilde{n_k}}|\tilde{n_k}\rangle\langle \tilde{n_k}|.
\eea
Therefore, the expected number of the particle creation is given by the Bose-Einstein form,
\bea\label{Planck}
\langle n_{\rm cr}\rangle= \frac{1}{{{e^{\frac{{{{\tilde \omega }_k}}}{{\tilde T}}}} - 1}} = |\beta_k|^2.
\eea
Eq. (\ref{Planck}) is a Planck spectrum and corresponds to a thermal state with following temperature,
\bea\label{TUnruh}
\tilde{T}=\frac{\tilde{\omega}_k}{2 \log \frac{1}{\tanh z_k}} =\frac{\tilde{\omega}_k}{\log (\frac{1}{\gamma})},
\eea
which yield to
\bea\label{T}
\hspace{ -.2cm}
\tilde{T}=-\tilde{\omega}_k \log^{-1} [ 1-\frac 1 {\pi^2 \mu^2 \csc {\rm h}^2 (\pi |\nu|) |J^\prime_\nu (\mu) J_{-\nu} (\mu)|^2}].
\eea
The used strategy, as stated above, has been applied to each quantum field theoretical scenario that is modelled by two-mode squeezing \cite{Carroll}. For example the well-known Unruh effect \cite{Unruh} or Hawking radiation in black hole scenario \cite{Davies}. In the Unruh effect, the vacuum state from view point of an inertial observer represents a thermal state, with the temperature $T\propto a$, for an uniformly accelerated observer with uniform acceleration $a$. In the radiation black hole scenario, near the horizon of the black hole, an inertial observer falling into the black hole by claiming that the field is in the vacuum state. Another observer that is accelerated and escapes the fall detects a thermal state. In this scenario the acceleration parameter is a function of the black hole mass. Also, both scenarios involve two-mode squeezing such that the temperature satisfies Eq. (\ref{TUnruh}) for the Unruh effect and similarly for the black hole case. In the both effects, the Unruh effect and the Hawking effect, by considering a asymptotically spatially flat metric, the initial vacuum state in new basis represents a canonical thermal state with temperature in Eq. (\ref{TUnruh}) which depends to scale factor. Accordingly, the used strategy in this paper is similar to other scenarios. A different scale factor for a asymptotically spatially flat spacetime leads to the initial vacuum state in the remote past represents a canonical thermal state in the remote future with the scale factor dependent temperature that is given by Eq. (\ref{TUnruh}).


Now, we can work out the entropy of particle creation using Eqs. (\ref{entropy}), (\ref{N}) and (\ref{T}),
\bea\label{PCE}
  S_{\rm cr}= \langle s\rangle =  \log \left(\frac {1+ \langle n_{\rm cr}\rangle} {\langle n_{\rm cr}\rangle}\right)^{\langle n_{\rm cr}\rangle},
\eea
where $\langle n_{\rm cr}\rangle$ is evaluated with respect to the cosmological and quantum field mode parameters by Eq. (\ref{ncr}).
\begin{figure*}
  \begin{minipage}[t]{.48\textwidth}
  \center
  \includegraphics[width=0.95\columnwidth]{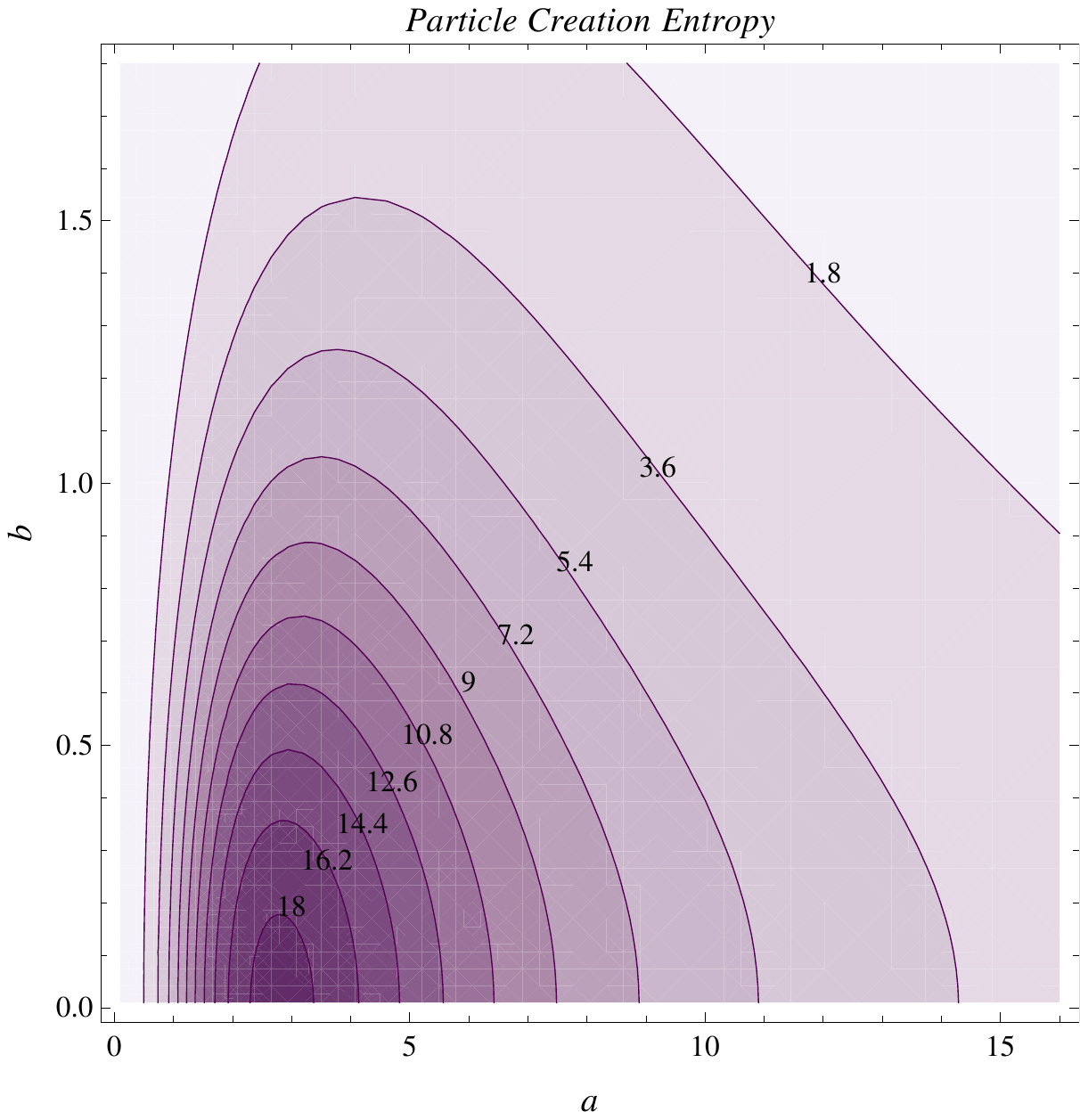}\\
  \caption{Contour plot of the Particle creation entropy with respect to the cosmic parameters $a,~b$ for $m=k=c=1$.}
  \label{Fig7}
\end{minipage}
  \begin{minipage}[t]{.48\textwidth}
  \center
  \includegraphics[width=0.95\columnwidth]{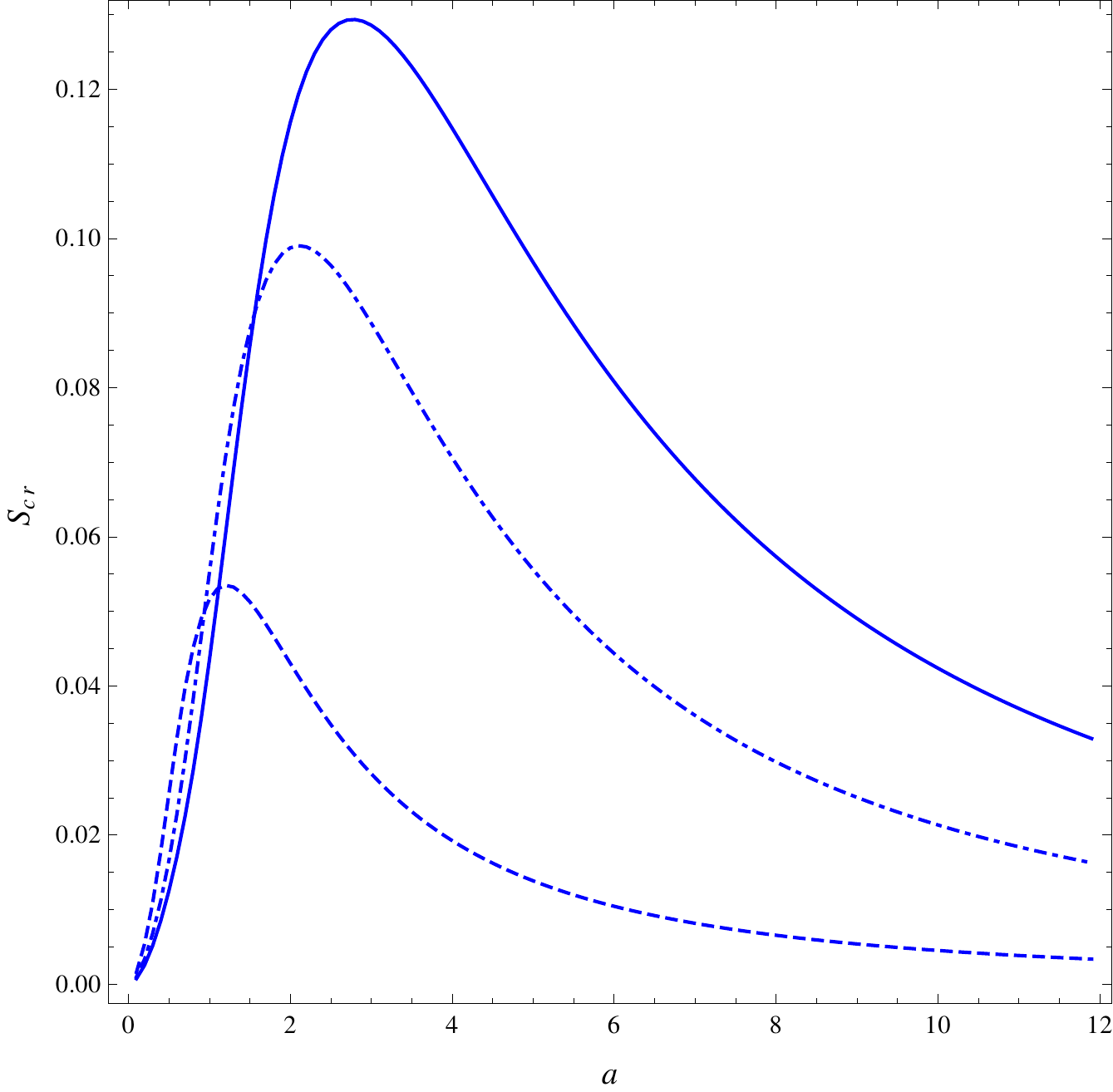}\\
  \caption{Particle creation entropy  with respect to the inflation rate $a$ for $b=0.01$ and for $m=k=c=1$, solid line, $m=k=c=0.8$, dotdashed line, and $m=k=c=0.5$, dashed line.}
  \label{Fig8}
  \end{minipage}
\end{figure*}
\section{entanglement entropy or particle creation entropy?}
\label{sec:en-cr}
Now we analyze entanglement and particle creation entropy. Substituting Eq. (\ref{ab}) and Eq. (\ref{alpha2}) in Eq. (\ref{beta2}) and Eq. (\ref{PCE}), entanglement entropy, $S_{\rm en}$, and particle creation entropy, $S_{\rm cr}$, are yielded.  $S_{\rm en}$ and  $S_{\rm cr}$ have been illustrated in terms of the field parameters, i. e. the mass and momentum of any modes, in Fig. \ref{Fig1} and Fig. \ref{Fig2}. The entropy quantities of massive scalar modes turn out to be a monotonic decreasing function of the momentum modes which the maximum value of them is occurred for the zero momentum modes. Both quantities with respect to the mass of the modes exhibit maxima at a same certain mass, $m_{\rm max}$, for each $k$ mode. The massless case has not entropy. Therefore it is clear that the $S_{\rm cr}$ behaves similar to the $S_{\rm en}$ in general.

It have to kept in mind that we have plotted Figs. \ref{Fig1} and \ref{Fig2} according to some assumptions so that the characteristics of the expanding universe, $a,~b,~c$, are fixed. In order to evaluation of the cosmological parameters, we depict every one of both entropies with respect to the mass and momentum of any modes separately. Both of $S_{\rm en}$ and $S_{\rm cr}$ are plotted in Figs.
\ref{Fig3} and \ref{Fig4} in terms of momentum of any modes for constant parameters $m,~a,$ and $b$ in Figs. \ref{Fig5} and \ref{Fig6} in terms of momentum of any modes for constant parameters $k,~a,~b$. In the cases being considered here, the quantities are plotted for four different values of the parameter of $c$.
    As Figs. \ref{Fig3} and \ref{Fig4} depict, entropy is decreasing with decreasing parameter $c$. Also Figs. \ref{Fig5} and \ref{Fig6} show that $m_{\rm max}$ is shifted to less positive values with increasing parameter $c$. It also gives rise to an increase in the value of entropy for this specific mass.
 It should be noted that both entropies, $S_{\rm en}$ and $S_{\rm cr}$, as well can be treated in the same way. Accordingly, let us only focus on the particle creation entropy in the following.

Doing a synchronous analysis of another cosmic parameters, $a,~b$, in Fig. \ref{Fig7} we can present the special peak that $S_{\rm cr}$ shows with respect to inflation parameter $a$. particle creation entropy exhibits maxima which their maximum value coincide with a special inflation rate, $a_{\rm max}$. $S_{\rm cr}$ as a function of inflation parameter $a$ is plotted in Fig. \ref{Fig8}. It shows particle creation entropy have maximum value for $a_{\rm max}$ due to fixed values $m,~k,~b,~c$. Thus we can get information from structure of the spacetime in which the field lives.

\section{further consideration}
\label{sec:further}
\begin{figure}
   \includegraphics[width=1\columnwidth]{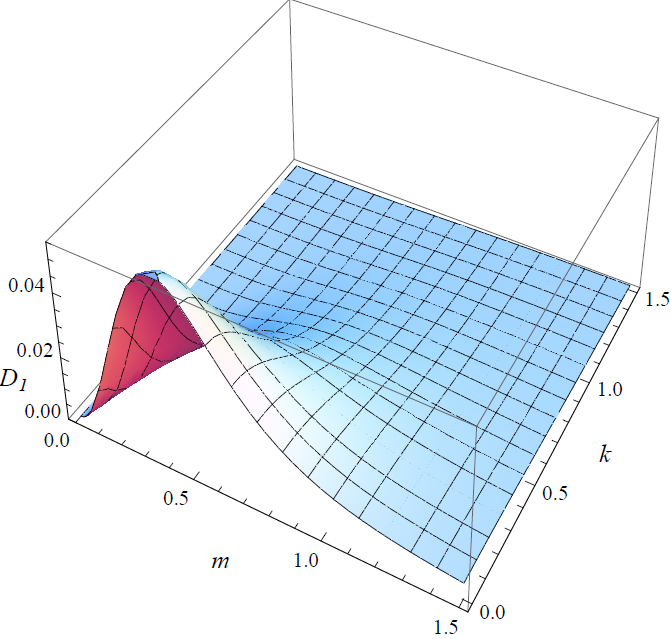}\\
  \caption{Difference between entanglement entropy and particle creation entropy, $D$, for Eq. (\ref{scale}), as a function of momentum $k$ and mass $m$ for fixed values $a=b=c=1$.
  }\label{D1}
\end{figure}
\begin{figure*}[!ht]
\begin{minipage}[t]{.48\textwidth}
  \center
  \includegraphics[width=1\columnwidth]{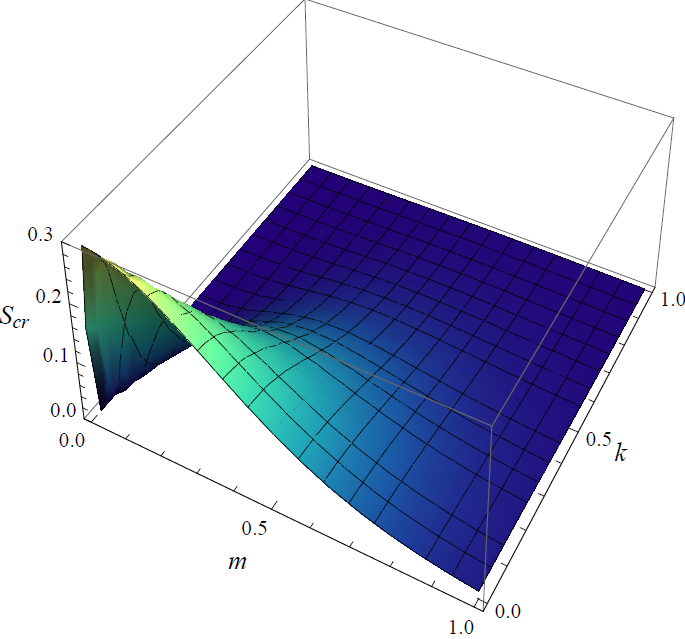}\\
  \caption{Particle creation entropy, $S_{\rm cr}$, for Eq. (\ref{scal2}), as a function of momentum $k$ and mass $m$ for fixed values $\epsilon=\rho=1$.
  }\label{Scr2}
  \end{minipage}
\begin{minipage}[t]{.48\textwidth}
  \center
  \includegraphics[width=1\columnwidth]{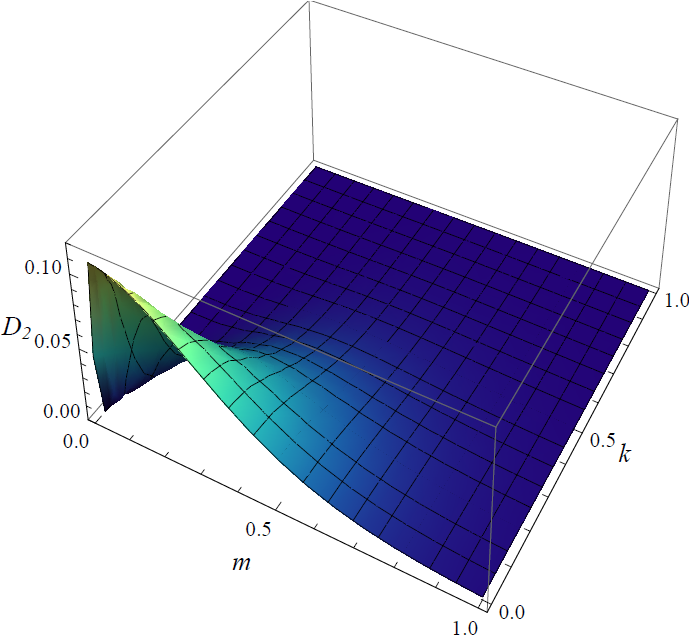}\\
  \caption{Difference between entanglement entropy and particle creation entropy, $D$, for Eq. (\ref{scal2}), as a function of momentum $k$ and mass $m$ for fixed values $\epsilon=\rho=1$.
  }\label{D2}
  \end{minipage}
\end{figure*}
More precisely, in this section we are going to survey the similarity of the expressions between entanglement entropy and particle creation entropy.
Using Eq. (\ref{N}), one can rewrite the entanglement entropy, Eq. (\ref{en}), in likeness with Eq. (\ref{entropy}) as follows
\begin{equation}\label{Sen}
     S_{\rm en}= \langle s\rangle =  \log \left(\frac{\left( {1+ \langle n_{\rm cr}\rangle} \right)^{\langle n_{\rm cr}\rangle+1}}{{\langle n_{\rm cr}\rangle}^{\langle n_{\rm cr}\rangle}} \right).
\end{equation}
Therefore, the relation between entanglement entropy, Eq. (\ref{Sen}) and particle creation entropy, Eq. (\ref{entropy}), is given by
\begin{equation}\label{Dif}
    S_{\rm en}=S_{\rm cr}+D,
\end{equation}
where
\begin{equation}\label{D}
   D=\log (1+\langle n_{\rm cr}\rangle).
\end{equation}
The difference between these two entropies, $D$, has been plotted in Fig. \ref{D1}, for the scale factor Eq. (\ref{scale}). $D$ is a function with respect to the number of particles. As it has been shown in Fig. \ref{D1}, the behavior of $D$ is similar to both entanglement entropy and particle creation entropy. Thus, $D$ scales the particle creation entropy with respect to the entanglement entropy that causes both quantities behave the same.
To understand this result is independent of choosing a scale factor, we control the difference between the two mentioned entropies above for a different scale factor \cite{Ball}, extremely applied in literature as follows
\begin{equation}\label{scal2}
    \Omega^2 ( \eta) = 1 + \epsilon (1+ \tanh ( \rho \eta)).
 \end{equation}
This scale factor presents a toy model of a universe which undergos a period of smooth expansion, where $\epsilon$ and $\rho$ are positive real
parameters indicating the total volume and rapidity of the expansion, respectively. The Robertson-Walker metric with Eq. (\ref{scal2}) is conformally flat at past and future infinity and the mentioned conditions in Sec. \ref{sec:exponent} admit also for such the scale factor. The Bogoliubov coefficients are such as follow
\begin{equation}\label{a}
   {\alpha _k} = {\left( {\frac{{{{\tilde \omega }_k}}}{{{\omega _k}}}} \right)^{\frac{1}{2}}}\frac{{\Gamma \left( {1 - \frac{{i{\omega _k}}}{\rho }} \right)\Gamma \left( { - \frac{{i{{\tilde \omega }_k}}}{\rho }} \right)}}{{\Gamma \left( { - \frac{{i\omega _k^ + }}{\rho }} \right)\Gamma \left( {1 - \frac{{i\omega _k^ + }}{\rho }} \right)}},
\end{equation}
\begin{equation}\label{b}
{\beta _k} = {\left( {\frac{{{{\tilde \omega }_k}}}{{{\omega _k}}}} \right)^{\frac{1}{2}}}\frac{{\Gamma \left( {1 - \frac{{i{\omega _k}}}{\rho }} \right)\Gamma \left( {\frac{{i{{\tilde \omega }_k}}}{\rho }} \right)}}{{\Gamma \left( {\frac{{i\omega _k^ - }}{\rho }} \right)\Gamma \left( {1 + \frac{{i\omega _k^ - }}{\rho }} \right)}},
\end{equation}
where $\omega_k ={\left( {{k^2} + {m^2}} \right)^{\frac{1}{2}}}$ for $\eta\rightarrow-\infty$, $\tilde \omega_k = {\left( {{k^2} + {m^2}\left[ {1 + 2\varepsilon } \right]} \right)^{\frac{1}{2}}}$ for $ \eta\rightarrow+\infty $ and $\omega _k^ \pm  = \frac{1}{2}\left( {{{\tilde \omega }_k} \pm {\omega _k}} \right)$.
 Therefore, by applying Eq. (\ref{N}), entanglement entropy, Eq. (\ref{Sen}), and particle creation entropy, Eq. (\ref{entropy}), are obtained. The entanglement entropy has been investigated in Ref. \cite{Fuentes1}. We have plotted the particle creation entropy in Fig. \ref{Scr2}. Also,
 difference between the two entropies, Eq. (\ref{D}), gives rise to Fig. \ref{D2}. In this case, as well as in other case, $D$ shows the similar behavior with both entanglement entropy and particle creation entropy and scales the particle creation entropy with respect to the entanglement entropy.
 As expected regardless of scaling, both Figs. \ref{D1} and \ref{D2} imply entanglement entropy shows the same treatment with respect to particle creation entropy.
%

It has been shown that quantum correlation between the systems as well as their thermal fluctuations can be used as resource to obey the thermodynamical work \cite{Funo}. Thus, the work can be extract from quantum entanglement. Now we study an addition operation to find a work quantity as an entanglement work, in analogy with Eq. (\ref{wfric}). In doing so, we apply the Eqs. (\ref{entropy}) and (\ref{T}), whereas the entropy is substituted by Eq. (\ref{en}). As a conclusion, the entanglement work done by unitary evolution of the spacetime is given by
\begin{equation}\label{wen}
\langle W \rangle_{\rm en} = \tilde{\omega} \left[ {\left\langle {{n_{\rm cr}}} \right\rangle  - \frac{{\log \left( {1 + \left\langle {{n_{\rm cr}}} \right\rangle } \right)}}{{\log \left( {\frac{{\left\langle {{n_{\rm cr}}} \right\rangle }}{{1 + \left\langle {{n_{\rm cr}}} \right\rangle }}} \right)}}} \right].
\end{equation}

In fact, We take advantage of a pure separable state as initial vacuum state in distant past. The new vacuum state will be an entangled state from the perspective of distant future observer. The von Neumann entropy of a pure separable state has zero value, indicating such a state can be completely described by its state function. Therefore, repeating observation of many copies of a prepared pure separable state will not provide any new information, which is not the case with a mixed state.
The quantum fluctuations of vacuum stemming from dynamical spacetime can be interpreted as the superposition of $| \tilde{n}_k \tilde{n}_{-k} \rangle $, from the view point of distant future observer. Thus the primary system will become necessarily an entangled state.
In the other words, the spacetime is doing the entanglement work, Eq. (\ref{wen}), onto the system and entropy is increased, which amount of entropy increasing is related to amount of entanglement.
In regarding with entanglement analysis of system is performed by the von-Neumann entropy, the initially separable state can be interpreted as an completely certain state, which there would not be any information due to it, corresponding to zero entropy.
Likewise, the dynamical spacetime depending to the selected metric will produce a entangled state with nonzero entropy.
In fact, the von Neumann entropy quantifies the mixedness of a state, or, if the state is seen as part of a pure entangled state, then it is the amount of entanglement in the total state.

On the other hand, The entropy production, as understood by Crooks \cite{Crooks}, is the difference between the probability distributions obtained in the forward and the reverse direction. The entropy production is proportional to inner friction and therefore, is directly proportional to particle creation. The forward process is considered as the expansion of spacetime beginning in an equilibriums state, describing by $H$. The contraction of this spacetime beginning in the the final adiabatic Hamiltonian $\tilde{H}_{ad}$ is the revers process \cite{Liu}.
In the other words, as we saw in sec. \ref{sec:entropy}, the thermodynamical work is done by spacetime and takes the initially ordered system away from thermal equilibrium. In this way, inner friction work creates the thermodynamic fluctuations and increases the entropy.
Therefore, from a physical viewpoint, both entropies have a physical concept. The source of produced entropy from both viewpoints is the dynamics of spacetime. This describes a interconnected relation between two approaches.
\section{Conclusion}
We studied particle creation from vacuum in a asymptotically spatially flat Robertson-Walker spacetime. Employing a scalar field in this spacetime, we used the past and future formalism to yield two sets of exact solutions for Klein-Gordon equation. These two sets are related to each other by a Bogoliubov transformation. Particulary, we selected an exponential scale factor and applied canonical method to evaluate the number of created particles. Investigation of entanglement was performed applying von Neumann entropy.

A two-mode squeezing model was introduced for describing the modes $k$ and $-k$ that was considered to interact unitarily during expanding.
The density matrix of two-mode squeezing system is represented a thermal state, so that the mode pairs are in thermal equilibrium with respect to each other in the classical spacetime background. Equilibrium temperature is dependent to the scale factor, but in a classical spacetime, quantum fluctuations of metric are not excluded, thus the classical spacetime is not affected by the created particles. The mode pairs and underling spacetime as a classical external source can be formed a closed quantum state driven out of equilibrium by unitary evolution. As a result the particles are created.

The procedure of particle creation gives rise to an increase of entropy which is proportional with inner friction. Accordingly, production entropy is exactly related to particle creation. The particle creation entropy was evaluated in terms of specific scale factor considered for the spacetime.
   Afterward, it was compared production entropy due to particle creation with von Neumann entropy as entanglement entropy. The comparison was shown that both of quantities can be treated in the same way, on equal footing, and both can be linked with each other and expressed as the increasing entropy of the universe due to particle creation.
Then using these quantities and dependence of them on the field parameters and cosmic parameters we can recover information from the underling structure of the spacetime.
Therefore, we can generalize formulation of thermodynamics to quantum formalism. Entanglement can be considered as a source of thermodynamics to properties associated with the universe. Our work implies that the thermodynamics properties of entanglement, in particular entropy, could really be coincided with the results found the customary quantum thermodynamics properties of the spacetime. As a consequents we can employ entanglement in the general thermodynamics picture of the universe. Indeed, it presents an important tool to study observable consequences in cosmology.
\acknowledgments
We specially appreciate Prof. V. Vedral for his comments. We thank
G. Najarbashi and M. N. Najafi for discussions.

\end{document}